\def\year{2022}\relax

\documentclass[letterpaper]{article} 
\usepackage{xcolor,soul}
\usepackage{aaai22}  
\usepackage{times}  
\usepackage{helvet}  
\usepackage{courier}  
\usepackage[hyphens]{url}  
\usepackage{graphicx} 
\usepackage{color}
\usepackage{natbib}  
\usepackage{caption} 
\DeclareCaptionStyle{ruled}{labelfont=normalfont,labelsep=colon,strut=off} 
\usepackage{algorithm}
\usepackage{algorithmic}
\usepackage{amsmath,amsthm,amssymb,amsfonts}
\usepackage{multirow}

\usepackage{newfloat}
\usepackage{listings}
\floatstyle{ruled}
\newfloat{listing}{tb}{lst}{}
\floatname{listing}{Listing}
\title{AOTree: Aspect Order Tree-based Model for Explainable Recommendation}
\author {
    Wenxin Zhao,\textsuperscript{\rm 1}
    Peng Zhang, \textsuperscript{\rm 1}
    Hansu Gu, \textsuperscript{\rm 2}
    Dongsheng Li, \textsuperscript{\rm 3}
    Tun Lu, \textsuperscript{\rm 1}
    Ning Gu \textsuperscript{\rm 1}
}
\affiliations {
    \textsuperscript{\rm 1} School of Computer Science, Fudan University, Shanghai, China\\
    \textsuperscript{\rm 2} Seattle, United States\\
    \textsuperscript{\rm 3} Microsoft Research Asia, Shanghai, China\\
    zhaowx21@m.fudan.edu.cn, 
    zhangpeng\_@fudan.edu.cn, 
    hansug@acm.org, 
    dongshengli@fudan.edu.cn, 
    lutun@fudan.edu.cn, 
    ninggu@fudan.edu.cn
}

\usepackage{bibentry}
\begin{document}

\maketitle

\begin{abstract}
Recent recommender systems aim to provide not only accurate recommendations but also explanations that help users understand them better. However, most existing explainable recommendations only consider the importance of content in reviews, such as words or aspects, and ignore the ordering relationship among them. This oversight neglects crucial ordering dimensions in the human decision-making process, leading to suboptimal performance. Therefore, in this paper, we propose Aspect Order Tree-based (AOTree) explainable recommendation method, inspired by the Order Effects Theory from cognitive and decision psychology, in order to capture the dependency relationships among decisive factors. We first validate the theory in the recommendation scenario by analyzing the reviews of the users. Then, according to the theory, the proposed AOTree expands the construction of the decision tree to capture aspect orders in users’ decision-making processes, and use attention mechanisms to make predictions based on the aspect orders. Extensive experiments demonstrate our method's effectiveness on rating predictions, and our approach aligns more consistently with the user’s decision-making process by displaying explanations in a particular order, thereby enhancing interpretability.

\end{abstract}
\section{Introduction}
\label{1 Introduction}
With the overwhelming amount of information on the Internet, recommender systems that aim to provide users with suitable items (products or services) by predicting a user's interest have been widely integrated into e-commerce, social network, and other web applications~\cite{karn2023customer, zheng2023automl}. Recommender systems play a crucial role in alleviating the information overload problem, thereby rendering the decision-making process more user-friendly and enhancing the overall user experience in domains such as online shopping, social interaction, and more. 
The dual objectives of recommender system development encompass not only accurately predicting user preferences but also enhancing user satisfaction and trust. This dual focus motivates research to emphasize both recommendation accuracy and user-centric metrics, with interpretability standing out as an important one. This involves creating explainable recommender systems capable of not just suggesting items but also providing users with explanations for those recommendations~\cite{tan2021counterfactual}. For instance, in an online shopping scenario, a recommendation for a mobile phone could include an explanation like \emph{``This phone is purchased by 80\% of your peers.''}


In the realm of explainable recommender systems, item reviews emerge as a valuable resource for generating explanations. Reviews offer insights from users' perspectives, serving a dual purpose: 1) a user's reviews contribute to characterizing their preferences~\cite{li2017neural}, and 2) an item's reviews can be leveraged to generate explicit explanations for its recommendation~\cite{2013Hidden}. So many recent studies in recommendations have focused on using advanced techniques like multi-view deep learning approaches to analyze features from user reviews as well as item reviews, and integrate them into recommender systems~\cite{elkahky2015multi, fan2022mv}. The key concept involves using attention mechanisms to capture the importance of different aspects reflected in user preferences and item characteristics from reviews and generating explanations accordingly~\cite{zhang2014users, pan2022accurate}. For instance, considering a user's emphasis on the aspect of price in their reviews, the explanation could be framed as \emph{``We recommend this phone to you because its price matches with your taste on affordability.''}


Despite the success of existing review-based explainable recommender systems, a crucial factor often overlooked is the aspect order in users' decision-making processes. The "Order Effects Theory" suggests that humans tend to follow a specific order of factors when making decisions~\cite{anderson1965primacy}, and the sequence in which factors are considered can influence decisions. For instance, a mobile photography enthusiast may prioritize the camera when buying a phone, while students might prioritize appearance and price over other factors. The different orders of considered aspects can significantly reveal users' preferences. Existing methods in explainable recommendations primarily focus on decisive factors in the decision-making process but neglect the ordering relationship between these factors.


Incorporating aspect order into review-based explainable recommendations poses key challenges. Firstly, constructing aspect order is intricate due to its personalized and dynamic nature for both users and items across diverse situations, presenting complexities in modeling. Users exhibit different decision orders, and likewise, various items entail distinct orders of consideration during selection. Furthermore, users' contemplation of each aspect is not predetermined but dynamically influenced by contextual factors such as item characteristics and prior decisions~\cite{meshram2016optimal}. Secondly, integrating aspect orders from both users and items for interpretable recommendations is non-trivial, given the inherent trade-off between explainability and effectiveness in optimization goals. Simultaneously optimizing these multiple aspects exacerbates the challenges in implementing and evaluating recommender systems.


To address these challenges, our focus is on introducing the effects of aspect order into explainable recommender systems. First, we validate the ``Order Effects Theory'' in the recommendation context by analyzing Yelp dataset\footnote{https://www.yelp.com/dataset}, delving into aspect orders embedded in users' decision-making sequences in online shopping. Second, to model and incorporate aspect orders, we propose an \textbf{A}spect \textbf{O}rder \textbf{Tree}-based (AOTree) explainable recommendation method. This three-stage process aims to enhance both recommendation accuracy and explanations. Specifically, we construct User-AOTree and Item-AOTree structures for users and items, respectively, capturing personalized and dynamic features of order effects. The final decisive sequence is derived by integrating aspect orders from both user and item perspectives, facilitating predictions that ensure recommendation accuracy, transparency, and explainability. Finally, extensive experiments validate our method's effectiveness, demonstrating higher accuracy compared to five state-of-the-art baseline methods. Additionally, our approach aligns more consistently with the user's decision-making process, thereby enhancing interpretability. To conclude, the main contributions of this work are summarized as follows:

\begin{itemize}
\item We validate the order effects in recommendation scenario by using a well-known recommendation dataset.

\item We propose an Aspect Order Tree-based (AOTree) explainable recommendation method by jointly characterizing order effects from the perspectives of both users and items.

\item Extensive experiments demonstrate our method's better accuracy and interpretability compared with several state-of-the-art methods.

\end{itemize}

The remainder of this work is organized as follows. First, we provide the related work to introduce the context of the current research. Then, we delve into data analysis to assess the existence of the theory and its applicability, outlining our motivation. Next, we elaborate on the AOTree explainable recommendation method framework. The subsequent section describes the experiments, along with the corresponding results and discussions. Finally, the conclusion is given.

\section{Related Works}
\label{2 Related Works}
Since our work is to apply the Order Effects Theory to the construction of explainable recommender systems, we divide the related work into two subsections: (a) \textit{Explainable Recommender Systems}; and (b) \textit{Order Effects Theory}.

\textbf{Explainable Recommender Systems.} 
In the domain of explainable recommender system research, there are two types of dominant models, namely model-agnostic and model-intrinsic approaches~\cite{lipton2018mythos}. Since there is generally a trade-off between performance and transparency in recommender systems, these two approaches have their own advantages and shortcomings regarding performance and transparency~\cite{zhang2020explainable}.

The model-agnostic approach, also named post-hoc explanation approach~\cite{peake2018explanation}, allows the recommendation model to be a black box and generates explanations after the recommendation results have been obtained. Aspect-aware techniques are commonly utilized to generate explanations in model-agnostic approach, i.e., explaining the recommendations with the corresponding aspects. \citet{zhang2014explicit} proposed an Explicit Factor Model (EFM) by aligning the latent factors with aspects, generating recommendations and explanations based on the matching degree of aspects between items and users. ~\citet{chen2016learning} extended EFM model to tensor factorization models and constructed the user-item-feature cube for pair-wise learning, which provides personalized recommendations as well as feature-based explanations. ~\citet{wang2018explainable} applied a joint factorization framework to integrate user preference and opinionated content for final recommendations and aspect-level explanations. Although model-agnostic methods can achieve high accuracy, they are complex and lack transparency. 

On the contrary, the model-intrinsic approach aims to develop an explainable model, i.e., interpretability is enhanced by improving the model's transparency~\cite{zhang2014explicit}. 
Decision tree is a basic transparent model, which characterizes each node by explicit criteria~\cite{kim2001application}. 
\citet{wang2018tem} proposed a tree-enhanced embedding model (TEM) for the explainable recommendation, combining the generalization ability of embedding-based models and the explainability of tree-based models. \citet{tao2019fact} introduced the FacT model to integrate regression trees to learn latent factors and use the learnt tree structure to explain recommendations. Specifically, the model predefined aspects to construct template-based explanations. \citet{bauman2017aspect} proposed the Sentiment Utility Logistic Model (SULM), which extracted aspects and the corresponding sentiments based on user reviews.
Besides, some works achieve interpretability by adopting a multi-rating approach to capture the relationship between various dimensions of criteria and the final ratings. ~\citet{zheng2017criteria} introduced "Criteria Chains" to establish the combination of criteria by contextual situations. ~\citet{fan2021predicting, fan2023improving} proposed a collective model to predict final ratings by jointly learning users' sub-scores over multi-criterion (service, locations, etc.), which could explain user preference features more naturally by presenting sub-scores of various criteria.
However, these models require more fine-grained information, such as the user sub-scores for multi-criterion. In addition, they impose higher demands on the accuracy of sub-scores, which means inaccuracy in any criterion will impact the final rating, resulting in overall poor performance~\cite{anwar2023efficient}.


\textbf{Order Effects Theory}. Order effects can be commonly observed and captured where information is presented as a sequence, such as the objects presented in psychology experiments~\cite{anderson1965primacy}, the evidence presented in court~\cite{maegherman2020law}, and the items presented in recommender systems~\cite{felfernig2007persuasive, zhao2021adversarial}. The Order Effects Theory claims that different sequence orders can cause different consequences~\cite{petty2001motivation}. It has been commonly used in research, and the studies concentrate on two topics: 1) how people perceive and process information in order to make decisions; and 2) how to present information in order to get people's acceptance.

For the first research topic, several experimental psychology studies investigate the influence of order effects in human memory~\cite{ebbinghaus2013memory}. The results show the importance of ordering in the decision-making process, and indicate that the first and the last items are easier to remember than the middle ones, recognized as Primacy and Recency~\cite{gershberg1994serial}. Also, ~\citet{beyond} mentioned a specific form of interpretability known as human simulability, and used sequences as the input of recurrent neural network (RNN) to construct a human-simulatable model, simulating the decision-making process of humans. For the second topic, \citet{maegherman2020law} demonstrated that the order of evidence presentation affected the cognitive dissonance and confirmation bias, which in turn affects the ability to persuade. \citet{petty2001motivation} carried out experiments to illustrate that the participants with different levels of motivation showed various susceptibility to order sequence, especially for the Primacy Effects. However, the inherent logical mechanism and mode of action of such decision-making models are not transparent.

Furthermore, recent researches have indicated that the recommendation process is a way of persuasion~\cite{gretzel2006persuasion}. 
Focusing on the influence of order effects for recommendation scenario, \citet{felfernig2007persuasive} stated that the serial position could influence users' degree of acceptance, especially for the information at the beginning and the end of the sequence~\cite{fogg1998persuasive,hitch1991prospective}. 
\citet{zhao2021adversarial} observed that users showed different interests for different orders of item sequence. Therefore, there could be an optimal order corresponding to the user's best acceptance. 

To conclude, although the Order Effects Theory has been investigated in several areas, the ordering relationship of attributes has not been considered in recommendations. To the best of our knowledge, this is the first work to capture order effects within the decision-making process in explainable recommender systems. By incorporating the Order Effects Theory into the construction of the model-intrinsic approach, we can enhance the approach's interpretability and also improve its performance.

\section{Preliminary Analysis}
\label{Data Analysis}
In this section, we explore the applicability of the Order Effects Theory within the realm of recommendation, analyzing the ordering relationship among user concerns in reviews. The analysis is conducted on the Yelp dataset, which is a publicly available user review dataset and commonly used for recommendation tasks. The statistics of the dataset after preprocessing are shown in Table~\ref{tab:Summary of datasets}. Following the sentiment analysis toolkit, named Sentires, built by ~\citet{zhang2014users}, the triplets (Aspect, Opinion, Sentiment) were extracted from the reviews to represent each datum, wherein Aspect is the focus in our analysis. 
This phrase-level sentiment analysis toolkit is a widely used tool for aspect analysis in recommendation system due to its accuracy and ease of use in capturing aspect and sentiment~\cite{zhang2020explainable, zhang2014explicit}. 
Based on Yelp dataset, we present a scenario where users make decisions regarding restaurants, highlighting the \textit{existence} of order effects and reveal its \textit{difference} among users/items. This comprehensive analysis yields valuable insights into the applicability of the Order Effects Theory. 

\begin{figure}[t]
    \begin{minipage}[t]{0.5\linewidth}
        \centering
        \includegraphics[width=\textwidth]{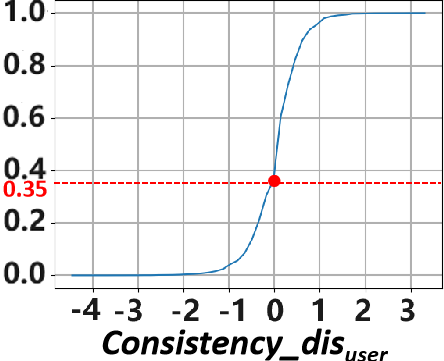}
        \centerline{(a) Among Users}
    \end{minipage}%
    \begin{minipage}[t]{0.5\linewidth}
        \centering
        \includegraphics[width=\textwidth]{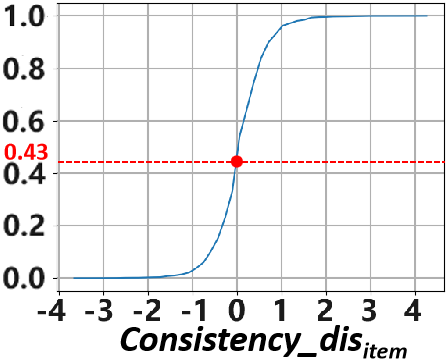}
        \centerline{(b) Among Items}
    \end{minipage}
    \vspace{-0.5em}
    \caption{The $Consistency\_dis_{user/item}$ value among users/items. The horizontal axis represents the proportion of users/items under specific consistency difference.}
    \label{fig:diff_ndcg}
    \vspace{-1.3em}
\end{figure}

We show the following randomly picked review from the Yelp dataset (by user \emph{Hie3\_7\_Nan3R3HaygoCFvA} for item \emph{VitNqJm8DIjw5D-Q-aiENQ}). For brevity, we display only key sentences containing the filtered extracted aspects (highlighted in bold). Aspect order can be considered as the order of the aspects that appeared as the sequence of the user's decision-making process~\cite{arapoff1967writing}.
\begin{quote}
\small
\textit{I've been here twice now, once for dinner and another time for the \textbf{lunch} buffet. The \textbf{food} is quite good, with nice hits of spice and savory \textbf{flavors}. The \textbf{prices} are average.}
\end{quote}
As shown in the review, the user initially emphasizes the aspect of \textbf{lunch} for mealtime, followed by considerations of the restaurant's \textbf{food} and \textbf{flavor} to evaluate taste. Subsequently, the aspect of \textbf{price} is addressed in later sentences, leading to the final experience step by step. The decisive chain can be identified as the aspect order: \{\textbf{lunch}, \textbf{food}, \textbf{flavor}, \textbf{price}\} (denoted as $O_{11}$). Similarly, another review for the same above user can be randomly picked and displayed as an aspect order of \{\textbf{lunch}, \textbf{food}, \textbf{price}\} (denoted as $O_{12}$). Further, for a different randomly selected user (user\_id is \emph{-\_FCaLa5eYXedOotc7J18Q}), one aspect order can be extracted as \{\textbf{food}, \textbf{service}, \textbf{food}, \textbf{chicken}, \textbf{service}, \textbf{table}, \textbf{spot}\} (denoted as $O_{21}$). It is obvious to see that the consistency of the two aspect orders for the same user ($O_{11}$ and $O_{12}$) is much higher than that for different users ($O_{11}$ and $O_{21}$), suggesting that the intra-consistency (within one user) of aspect order is higher than the inter-consistency (between different users).

We further aim to generalize the above findings by randomly selecting 10,000 users from Yelp and then constructing 10,000 pairs of intra-user reviews (two reviews from the same user) and 10,000 pairs of inter-user reviews (two reviews from two different users). For each pair of intra-user or inter-user reviews, we utilize the NDCG metric~\cite{kanoulas2009empirical} (the details of NDCG are described in Experiment Section) to evaluate the consistency of aspect order between the two reviews (named $intra\_cons_{user}$ or $inter\_cons_{user}$, respectively). To highlight the difference between a user's $intra\_cons_{user}$ and $inter\_cons_{user}$, we construct a measurement named $consistency\_dis_{user}$ which is computed as $intra\_cons_{user}$ minus $inter\_cons_{user}$. It can reflect whether a user has a specific order preference. The CDF (Cumulative Distribution Function) of $consistency\_dis_{user}$ among the 10,000 users is shown Figure~\ref{fig:diff_ndcg} (a). From the figure, we can see the $intra\_cons_{user}$ of nearly $70\%$ users is greater than their corresponding $inter\_cons_{user}$ ($p < 0.001$), indicating that most users have specific aspect order considerations. We can define such users as \textit{Sensitive Users}. Moreover, it suggests that a proportion of users are \textit{Strong Sensitive Users}, i.e., the $intra\_cons_{user}$ is much higher than the corresponding $inter\_cons_{user}$. For example, about $20\%$ of users' $consistency\_dis_{user}$ values are higher than $0.5$ ($p < 0.001$). These results confirm our above observation, i.e., for most users, the $intra\_cons_{user}$ is overall higher than the corresponding $inter\_cons_{user}$, suggesting that a user tend to have aspect order preference while different users exhibit some differences.

Similarly, we conduct an analysis from the perspective of items, i.e., randomly selecting 10,000 items and constructing 10,000 pairs of intra-item reviews (two reviews from the same item) and 10,000 pairs of inter-user reviews (two reviews from two different items). After that, $intra\_cons_{item}$ (the consistency of aspect order between two intra-item reviews), $inter\_cons_{item}$ (the consistency of aspect order between two inter-item reviews), and $consistency\_dis_{item}$ ($intra\_cons_{item}$ minus $intra\_cons_{item}$) are evaluated. The result is shown in Figure~\ref{fig:diff_ndcg} (b). It shows that for over $60\%$ items, $intra\_cons_{item}$ is overall higher than the corresponding $inter\_cons_{item}$, suggesting that an item tends to have a specific aspect order considered by users and different items have some differences ($p < 0.001$). Additionally, approximately $20\%$ items are strong sensitive with the $consistency\_dis_{item}$ value higher than $0.5$ ($p < 0.001$).

\begin{figure}[]
  \centering
  \includegraphics[width=0.95\linewidth]{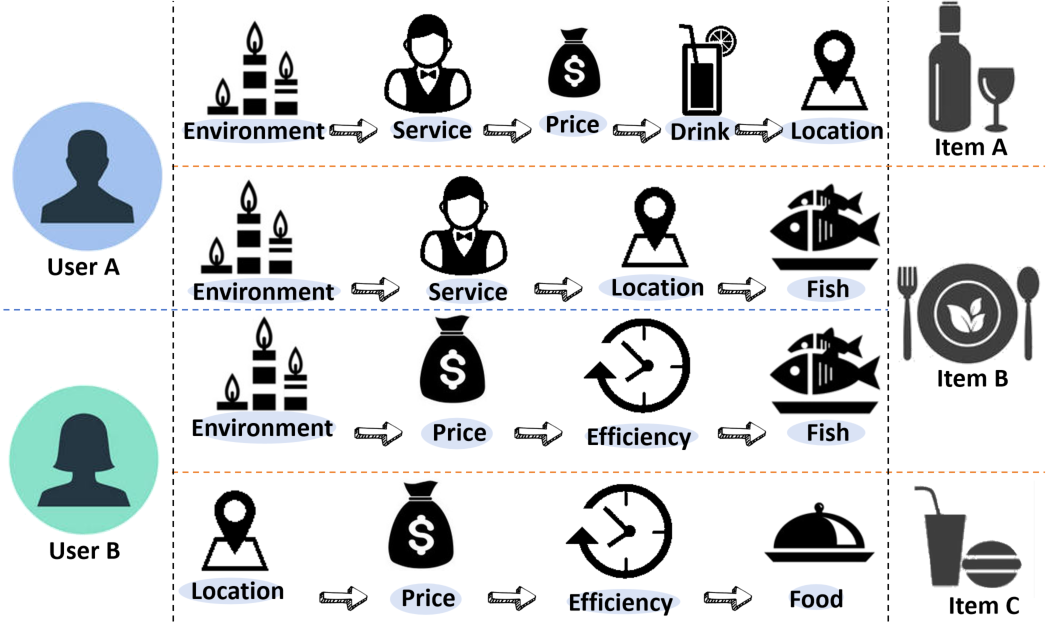}
  \caption{Example of the existence and variances of order effects in reviews.}
  \label{fig:example_preliminary}
  \vspace{-1.3em}
\end{figure}

Based on the above analysis, we confirm the existence of order effects in user reviews and their variances. Specifically, each user prefers an overall specific aspect order, and each item also tends to have an overall specific aspect order considered by users (Existence). However, a user's aspect order preference can vary across different items to a certain extent, and an item's aspect order considered by different users can also vary to some degree (Variance). We use a real-life example to illustrate the conclusion more intuitively (See Figure~\ref{fig:example_preliminary}). When user $A$ is selecting a restaurant for himself, he generally tends to prioritize \textbf{Environment} and \textbf{Service}, while other aspects like \textbf{Location} and \textbf{Price} are considered occasionally, which indicates his overall aspect order preference and some variance across items. Similarly, an item like item $B$ tends to be judged from the perspective of \textbf{Environment} due to its remarkable dining environment. Beside that, different users may evaluate item $B$ in different aspect order according to own aspect preference. It suggests item $B$'s overall specific aspect order considered by users and the variance across users.

\vspace{-0.5em}
\section{Methodology}
\label{4 Methodology}
\begin{figure}[]
  \centering
  \includegraphics[width=0.95\linewidth]{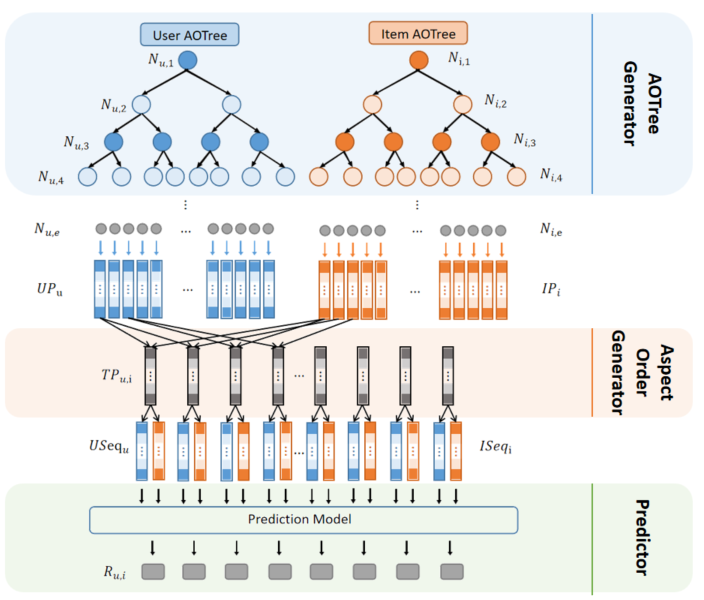}
  \caption{The algorithmic framework of AOTree. 
  }
  \label{fig:overview}
  \vspace{-0.5cm}
\end{figure}

In this section, we introduce our proposed Aspect Order Tree-based explainable recommendation method --- AOTree, which aims to simulate human decisive behavior according to Order Effects Theory. 
We first present the Problem Formulation of this work followed by the general architecture shown in Figure~\ref{fig:overview}, which consists of three key modules: 1) the AOTree Generator, 2) the Aspect Order Generator, and 3) the final Predictor.
\vspace{-0.5em}
\subsection{Problem Formulation}
\label{4_1 Problem Formulation}

In this work, we target at the {\bf rating prediction} task.
Let $\mathcal{U}=\{u_1, u_2, ..., u_m\}$, $\mathcal{V}=\{v_1, v_2, ..., v_n\}$ denote the user set and item set, respectively, where $m$ and $n$ represent the number of users and items. 
Then, $r_{i,j}$ is the rating value for a user $u_i$ toward an item $v_j$. 
And a set of observed data is represented as $\mathcal{O} = \{(u, v)|u \in \mathcal{U}, v \in \mathcal{V}, u\ \textit{has reviewed}\ v\}$.

The construction of the sentiment set, represented by triplets (Aspect, Opinion, Sentiment), follows the same procedure outlined in the Data Analysis Section. Suppose we finally get the aspect set $\mathcal{A}=\{a_1, a_2, ..., a_l\}$, where $l$ denotes the number of aspects. 
Then, we construct the user aspect matrix $X \in \mathbb{R}^{m*l}$ as well as the item aspect matrix $Y \in \mathbb{R}^{n*l}$ as~\cite{zhang2014explicit}:
\vspace{-0.5em}
\begin{small}
\begin{equation}
\label{Equ: aspect matrix}
\begin{split}
X_{i,k}&=\begin{cases}
   0, ~\text{if } u_i\ \text{did  not mention aspect}\ a_k,\  \\
   1+(N-1)(\frac{2}{1+exp(f_{-i,k})}-1),~\text{otherwise.}
\end{cases} \\
Y_{j,k}&=\begin{cases}
   0, ~\text{if  aspect}\ a_k\ \text{was\ not\ mentioned\ for\ item}\ v_j,\\
   1+\frac{(N-1)}{1+exp(-f_{j,k} \cdot s_{j,k})},~\text{otherwise.}
\end{cases}\nonumber
\end{split}
\end{equation}
\end{small}
\vspace{-0.5em}

$N$ denotes the rating scale, which is 5 in our datasets. $f_{j, k}$ is the frequency of user $u_i$ mentioning aspect $a_k$, and $s_{j, k}$ is the average sentiment calculated from the sentiment set. Since $X_{i,k}$ and $Y_{j,k}$ measure the importance of aspect $a_k$ to user $u_i$ and item $i_j$, respectively, we thus call them ``aspect importance'' in this paper and $X$ and $Y$ are called {\bf individual aspect importance matrices}.

\vspace{-0.5em}
\subsection{Group Aspect}
\label{4_1 Group Aspect}
Before building AOTree, we first process the user/item aspect matrix into a general representation for all users/items rather than construct trees for each user/item in order to reduce the time complexity. 
We take the user AOTree generation as an example, the process of which is the same as item AOTree generation. 

However, getting the mean value of each aspect according to individual user aspect importance matrix $X$ ignores the portion of contribution for each user because the size and effectiveness of each review are different among users. 
Inspired by the weighted technique~\cite{chen2018neural}, we integrate the aspect matrix by considering the contribution of each user, which means weighing the user preference value for each aspect. 
The formulation can be described as follows:
\begin{small}
\begin{equation}
\label{Equ: group user aspect opt}
    \hat X_{k} = \sum_{i=1}^m X_{i,k} * W_i,\quad
    W_i = \frac{N_i}{\sum_{i=1}^m N_i},
\end{equation}
\end{small}

where $\hat X_k$ represents the general (globally weighted) aspect importance on aspect $a_k$ over all users ($\hat X$ is called {\bf general aspect importance vector}). And $W_i$ is the portion of contribution from user $u_i$, calculated by $N_i$, the number of reviews written by user $u_i$. 
We could finally get the general user aspect importance vector $\hat X=\{\hat X_1, \hat X_2, ..., \hat X_l\}$ for all aspects. Similarly, the general item aspect importance vector $\hat Y=\{\hat Y_1, \hat Y_2, ..., \hat Y_l\}$ can be obtained.
\vspace{-0.5em}
\subsection{AOTree Generator}
\label{4_2 AOTree Generator}

The tree-based model is considered to be explainable and transparent when applied to recommendation tasks~\cite{zhang2020explainable}. Also, it is easy for humans to simulate and thus understand and trust~\cite{beyond}.
Rather than learning features for prediction like other deep learning methods, the tree-based model learns decision rules from datasets. 
However, the traditional decision tree method aims to solve classification problems, which is not applicable to our situation. 
So, we design our own decision rules to split items/users according to the matching degree within aspect quality/preference value to find an aspect order representing the considering decision process of users. 

Due to the variances of aspect order effects suggested in the Preliminary Analysis Section, we personalize the order construction for both users and items when building the AOTree, respectively.
To be more specific, when building the User-AOTree, we could split user sets and get the considering aspect sequence for each user by matching the general item aspect importance vector $\hat Y$ with the individual user aspect importance matrix $X$, meaning how the sequence of general item quality is shown according to different users. 
The Item-AOTree can be constructed in the same way with the corresponding general user aspect importance vector $\hat X$ with the individual item aspect importance matrix $Y$. 

In general, inspired by the decisive process of Markov Decision Processes (MDP), the goal is to maximize its reward stream, and thus, each decisive step is chosen based on the best current state~\cite{shani2005mdp}. There are various approaches existing for obtaining optimal policy, exactly or approximately. In our scenario, each step of our aspect order is defined as the chosen aspect, which is based on the minimal Split Expense for the optimal Split Value.
Specifically, our AOTree is constructed as the following three steps:

\begin{enumerate}
\item Calculate Split Value (SV) for each aspect, representing the optimal split criteria to match user preference and item characteristic corresponding to each aspect; 
\item Calculate Split Expense (SE) for each aspect with its corresponding SV, representing the reward for each aspect; 
\item Choose aspect and corresponding SV with the minimal SE value, representing the optimal decision step. 
\end{enumerate}
For each node, we repeat the above three steps to split sets into different nodes until the node contains only one element, recognized as the leaf. 
However, in order to improve efficiency, we limit the depth of the tree as a pre-defined threshold, meaning that the leaf node may contain more than one element but with very similar characteristics.

\subsubsection{Calculate Split Value (SV)}
\label{4_2_1 SV}
We first traverse all aspects to calculate the split value for each aspect, representing the matching degree between the user's preference and the item's quality. 
Rather than directly using values in the general item aspect importance vector ($\hat Y$) to split nodes, which may cause biases among different users, we focus on the degree of ranking matching between users and items on each aspect~\cite{nevo1989validation}.

In detail, for each aspect, we first get the item quality rank position ($PI_k=1, 2, ..., l$) according to the general item aspect importance vector ($\hat Y$) and find the corresponding user rank position ($PU_k$) in the individual user aspect importance matrix ($X$) for the same aspect ($a_k$). 
$PI_k$ and $PU_k$ satisfy the Equation ~\ref{Equ: Calculate Split Value_Equation}, which means the item's quality rank matches the user's preference rank.
\begin{small}
\begin{equation}
\label{Equ: Calculate Split Value_Equation}
    PI_k/l = PU_k/m.
\end{equation}
\end{small}
As the values of $PI_k$, $m$ and $l$ are already known, the corresponding user rank position ($PU_k$) could be calculated.

Then, the Split Value (SV), the value in the individual user aspect importance matrix ($X$) at the corresponding position, could be obtained by the following equation inspired by the matching-ranking technique~\cite{nevo1989validation}:
\begin{small}
\begin{equation}
\label{Equ: Calculate Split Value}
    \frac{PU_k-PU_{kl}}{PU_{kr}-PU_{kl}} = \frac{SV_{k}-X_{P_{kl},k}}{X_{PU_{kr},k}-X_{PU_{kl},k}}.
\end{equation}
\end{small}
As $PU_{k}$ is obtained as a decimal, $PU_{kr}$ and $PU_{kl}$ represent the right and the left integer rank position index of $PU_k$ ($PU_{kr} > PU_{k} > PU_{kl}$). 
And $X_{PU_{kr},k}$ and $X_{PU_{kl},k}$ are the corresponding values in the individual user aspect importance matrix ($X$) for aspect $a_k$. As $SV_k$ is the only unknown value in Equation~\ref{Equ: Calculate Split Value}, it can be easily derived given other values.
\begin{figure}[t]
  \centering
  \includegraphics[width=\linewidth]{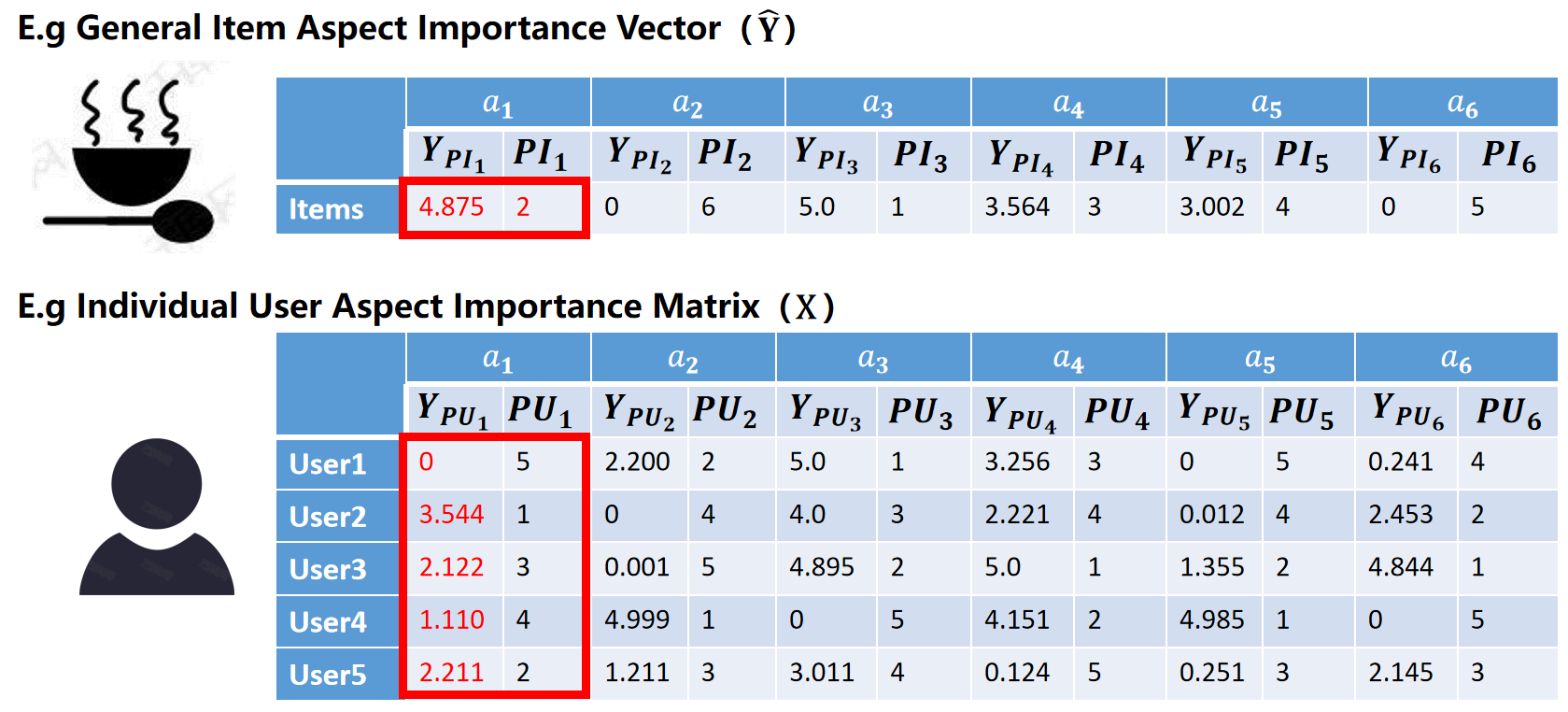}
  \vspace{-0.7cm}
  \caption{Example of calculating Split Value (SV).}
  \label{fig:split_value_eg}
  \vspace{-1.3em}
\end{figure}

The process can be illustrated as a specific example in Figure~\ref{fig:split_value_eg} for a better understanding. 
Take aspect $a_1$ as one of the traversals, so the $k$ is 1 in this example. Then, the value for $PI_1$, $l$ and $m$ are 2 (item quality rank position for $a_1$), 6 (number of aspects) and 5 (number of users), and we can obtain $PU_1=m * PI_1/l=5 * 2/6=1.667$ (according to Equation~\ref{Equ: Calculate Split Value_Equation}). 
The corresponding value for $PU_{1r}$, $PU_{1l}$ are 2 and 1, then $X_{PU_{1r},k}$ and $X_{PU_{1l},k}$ are $3.544$ and $2.211$, respectively corresponding to $User2$ and $User5$. And the final $SV$ value for aspect $a_1$ is $SV_1 = \frac{(1.667-1)}{(2-1)}*(3.544-2.211)+2.211=4.433$ (according to Equation~\ref{Equ: Calculate Split Value}). 
We could get $SV$ value for all aspects in the similar way.
\subsubsection{Calculate Split Expense (SE)}
\label{4_2_2 SE}
Then, we decide which aspect should be chosen as the node split criterion and split users by comparing the corresponding aspect value in the individual user aspect importance matrix ($X$) with $SV$ based on the chosen aspect. Take $a_k$ as an example, if the aspect importance value for the user ($X_{ik}$) is larger than $SV_k$, the user is split into the right node, otherwise, into the left node.
Our goal is to find one aspect that could make the best matching between user preference and item quality, simultaneously ensuring the robustness of the decision tree.
Targeting the rationality and robustness of the decision tree construction, we specify the split principles into the following three components, $SE_1$, $SE_2$ and $SE_3$ aiming at the child nodes after splitting the node. The specific division principles refer to classification and clustering methods, such as Support Vector Machine (SVM) and K-Nearest Neighbors (KNN), aiming to enhance intra-class similarity while promoting inter-class dissimilarity: (suppose $a_k$ is the split criterion for the current node):

1) $SE_1$: users in the same child node should be similar on aspect $a_k$, i.e., with minimal distance:
\begin{small}
\begin{equation}
\textstyle
\label{Equ: Split_Principle_1}
\min SE_1=\sum_{i=1}^m|X_{ik}-\bar{X}_{k}|,
\end{equation}
\end{small}
where $\bar{X}_{k}$ and $X_{ik}$ are the average aspect importance value in the individual user aspect importance matrix ($X$) and the specific value for user $i$, both on aspect $a_k$.

2) $SE_2$: users in the different child node groups should be different on aspect $a_k$, i.e., with maximal distance:
\begin{small}
\begin{equation}
\label{Equ: Split_Principle_3}
\max SE_2=|\bar{X}_{right,k}-\bar{X}_{left,k}|,
\end{equation}
\end{small}
where $\bar{X}_{right,k}$ and $\bar{X}_{left,k}$ are the average aspect importance values for aspect $a_k$ in the right and left child node.

3) $SE_3$: users in the same child node group should be different on aspects except $a_k$, to continually split users:
\begin{small}
\begin{equation}
\textstyle
\label{Equ: Split_Principle_4}
    \max SE_3=\sum_{o=1}^{l-k}\sum_{i=1}^m|X_{io}-\bar{X}_{o}|,
\end{equation}
\end{small}
where $X_{io}$ is the aspect importance for user $i$ on aspect except $a_k$, and $\bar{X}_{o}$ is the average aspect importance for all users on aspect $a_o$.

Then, we could get the Split Expense (SE) value by traverse all aspects according to the corresponding Split Value (SV) as follows:
\begin{small}
\begin{equation}
\label{Equ: Split Expense}
    SE = NV * SE_l*SE_r,
\end{equation}
\end{small}
where $NV$ represents the normalization constant to normalize the value of $SE$, and we set it to $10^{N_r*N_l}$ in our method, where $N_r$ and $N_l$ are the number of users in each child node. 
The subscripts ``$r$'' and ``$l$'' denote the values of the right and left child nodes, respectively.
$SE_l$ and $SE_r$ are Split Expense (SE) for the left and right nodes after splitting, which could be obtained by: 
\begin{small}
\begin{equation}
\label{Equ: Split Expense_2}
    SE_l = \frac{SE_{l1}}{SE_{l2}*SE_{l3}}, \quad 
    SE_r = \frac{SE_{r1}}{SE_{r2}*SE_{r3}}.
\end{equation}
\end{small}
\subsubsection{Choose Aspect and Split Value}
\label{4_2_3 Choose Aspect}
We finally choose the aspect with the minimal Split Expense (SE) in order to satisfy our goal:
\begin{small}
\begin{equation}
\label{Equ: Choose Aspect}
    k = \mathop{\arg\min_k} \ \  SE.
\end{equation}
\end{small}
Then the aspect $a_k$ is the split aspect for this node, and the corresponding Split Value (SV) can be obtained by Equation~\ref{Equ: Calculate Split Value}.

\subsubsection{Build AOTree}
\label{4_2_4 Build AOTree}
We continue the above three steps to decide on each node with a split aspect and its corresponding Split Value (SV). 
The criteria for stopping the division is until each node has only one item or the tree depth has reached the predefined threshold.

The obtained User-AOTree represents how the sequence of general item quality is shown according to different users. 
By deciding the split aspect for each node, we could construct aspect order for certain user sets.
In each leaf node, users with similar aspect quality share the same aspect order (e.g., $UP_i=\{N_1, N_2, ..., N_e\}$ for user $u_i$, where $N$ and $e$ denote the aspect id and the path length).

Similarly, the Item-AOTree can be built following the same process (e.g., the aspect order is $IP_j=\{N_1, N_2, ..., N_e\}$ for user $v_j$), representing how the sequence of general user preference is shown according to different items.
\vspace{-1em}
\subsection{Aspect Order Generator}
\label{4_3 Aspect Sequence Generator}
\vspace{-0.5em}
For each interaction (user-item pair), we could convert it into aspect order $UP_i$ for $u_i$ and aspect order $IP_j$ for $v_j$ according to the User-AOTree and Item-AOTree, respectively. 
The above two orders are considered separately from the user and item sides, providing information for the final decision sequence. 
So we combine these two considering orders by a simple average:
\begin{small}
\begin{equation}
\label{Equ: Combine Sequence}
    Ind_{k} = \frac{Ind_{UP, k} + Ind_{IP, k}}{2},
\end{equation}
\vspace{-0.5em}
\end{small}
where $Ind_{UP, k}$ and $Ind_{IP, k}$ denote the ranking index of aspect $a_k$ in $UP_i$ and $IP_j$, respectively. 
Then, the average aspect index $Ind_{k}$ can be sorted to construct the final aspect sequence $TP = \{TP_{1}, TP_{2}, ...,  TP_{e}\}$, and each element is the id for aspect, representing the decisive aspect process for certain user toward certain item. 
We fill each sequence into a fixed-length $e$ with random aspect id if the length is less than $e$.

Corresponding to the aspect id sequence $TP$, we could get the final user or item aspect importance sequence $USeq$ or $ISeq$ by picking the value of the corresponding aspect id from the original individual user/item aspect importance matrix ($X$/$Y$). For example, the value for $USeq = \{USeq_{1}, USeq_{2}, ...,  USeq_{e}\}$ is picked from $X_i$ based on the aspect id in $TP$, which means, $USeq_{i}$ is equal to $X_{i, TP_{i}}$.
\vspace{-0.8em}
\subsection{Prediction with Aspect Order}
\label{4_4 Prediction with Aspect Sequence}
\vspace{-0.5em}
The final rating prediction result is obtained mainly by two components, the decision process information, which is the core part of our method, and the user/item context information. 
In the learning process, we refer to the traditional latent factor method, but extend with our designed components: applying position embedding and self-attention layer to the sequence, and then the final ratings are obtained with a linear model.
\vspace{-0.8em}
\subsubsection{Position Embedding Layer}
\label{4_4_1 Position Embedding Layer}
Given the aspect order generated by AOTree ($TP$), which is represented by aspect id, we project each id into an embedding vector with size $d$. Then, we could obtain the path embedding $\hat {TP} = \{\hat {TP_{1}}, \hat {TP_{2}}, ..., \hat {TP_{e}}\}\in \mathbb{R}^{e \times d}$.

In order to highlight the position of each aspect in the sequence, we consider combining our sequence with a position embedding vector. 
Inspired by \citet{kang2018self}, we add a position embedding $E=\{E_1, E_2, ..., E_e\}\in \mathbb{R}^{e \times d}$ to the path embedding $\hat {TP}$ to get $\mathcal{N}$, described as:
\begin{small}
\begin{equation}
\label{Equ:Position Embedding}
    \mathcal{N}=
    \left[
    \begin{array}{c}
    \hat {TP_{1}} + E_{1}  \\
    \hat {TP_{2}} + E_{2}  \\
    ...\\
    \hat{TP_{e}} + E_{e}
 \end{array}
 \right].
\end{equation}
\end{small}

\subsubsection{Self-Attention Layer}
\label{4_4_2 Self-Attention}
We use scaled dot-product attention to achieve self-attention as~\cite{vaswani2017attention}:
\begin{small}
\vspace{-0.5em}
\begin{equation}
\label{Equ:attention}
    Attention(\textbf{Q}, \textbf{K}, \textbf{V}) = softmax(\frac{\textbf{Q}\textbf{K}^T}{\sqrt{d}}) \textbf{V},
\end{equation}
\end{small}
where $\textbf{Q}$, $\textbf{K}$ and $\textbf{V}$ denote queries, keys and values, respectively. 
The scaling factor $\sqrt{d}$ is to avoid overly large values of the inner product, especially when the dimensionality is high, and $\textbf{K}$ and $\textbf{V}$ are with the same values. 
In our method, $\mathcal{N}$ is converted to the three input for self-attention as:
\begin{small}
\begin{equation}
\label{Equ: Att}
    Att = Attention(\mathcal{N}\textbf{W}^{Q}, \mathcal{N}\textbf{W}^{K},\mathcal{N}\textbf{W}^{V}),
\end{equation}
\end{small}
where $\textbf{W}^{Q}$, $\textbf{W}^{K}$ and $\textbf{W}^{V}$ are projection matrices. 
Then, the result is added to the normalized input $\mathcal{N}$ to obtained the Sequence Feature (SF) as:
\begin{small}
\begin{equation}
\label{Equ:Added Att}
    SF = LayerNorm(\mathcal{N} + Att).
\end{equation}
\vspace{-0.8em}
\end{small}

Due to the order consideration, we should mask the first $t-1$ aspects in the order when processing the $t$-th layer of aspect, so we add a triangular matrix for masking the unseen position. 
Layer normalization is used to normalize the inputs, which is beneficial for stabilizing and accelerating neural network training, represented as:
\begin{small}
\vspace{-0.5em}
\begin{equation}
    LayerNorm(x) = \alpha \odot \frac{x - \mu}{\sqrt{\sigma^2+\epsilon}} + \beta,
\end{equation}
\end{small}
where $\mu$ and $\sigma$ represent mean and variance of $\mathcal{N}$, $\alpha$ is scaling factor, $\beta$ is bias term, and $\odot$ is element-wise product. 

Finally, we optimize aspect order for user ($USeq$) and item ($ISeq$) by element-wise product with SF as:
\begin{small}
\begin{equation}
\label{Equ: sequence with feature}
\begin{split}
    \hat USeq = USeq \odot SF.\\
    \hat ISeq = ISeq \odot SF.
\end{split}
\end{equation}
\end{small}
\begin{table*}[htb!]\small
\vspace{-0.5cm}
\caption{The statistics of the five datasets used in the experiments.}
\vspace{-0.3cm}
\label{tab:Summary of datasets}
\centering
\vspace{-0.3em}
\begin{tabular}{c|lllllll}
\hline
Dataset             & \#Aspect          & \#Users & \#Items & \#Reviews & T\_u/T\_i & T\_a &  Density \\ \hline

Cells Phones and Accessories & 1,016 & 14,782   & 10,825   & 109,956 & 20 & 20 & 44.80  \\
Office Product     &433          & 2,433    & 2,074    & 27,091  & 20 & 20 &44.63\\
Patio, Lawn and Garden        & 399      & 1,137    & 1,103    & 9,926  & 20 & 20 & 41.48  \\ 
Digital Music     &215          & 4,392    & 3,247    & 44,595  & 5 & 20 &32.34\\ 
Yelp  &  28 & 19,225   &  13,083   & 81,777 & 3 & 5 & 5.05 \\ \hline
\end{tabular}
\end{table*}
\vspace{-0.8em}
\subsubsection{Prediction Layer}
\label{4_4_3 Prediction Layer}

In the prediction layer, we apply a simple linear model to integrate all the components:
\begin{small}
\begin{equation}
\small
\label{Equ: Rating Prediction}
    \hat r_{ij}=W_1^T(\hat USeq_i\odot \hat ISeq_j)+W_2^T(p_i\odot q_j)+b_u+b_i+\mu,
\end{equation}
\end{small}
where $W_1 \in \mathbb{R}^d$ and $W_2 \in \mathbb{R}^d$ are weight matrices of the linear model, and $p_i$ and $q_j$ denote the ID embedding for user $u_i$ and item $v_j$. 
$\hat USeq_i\odot \hat ISeq_j$ is the decision process information, and $p_i\odot q_j$ represents the interaction between user $u_i$ and item $v_j$.

\vspace{-0.5em}
\subsection{Model Training}
\label{4_5 Model Learning}
We use Mean Square Error (MSE) as our objective function: 
\begin{small}
\begin{equation}
\label{mse}
\textstyle
\textbf{MSE}=\frac{1}{N}\sum_{(i,j)\in \mathcal{O}}(r_{ij}-\hat{r}_{ij})^2,
\end{equation}
\end{small}
where $r_{ij}$ denotes the ground truth rating for user $u_i$ on item $v_j$, and $\hat {r}_{ij}$ is the predicted rating from our method.

\vspace{-0.5em}
\section{Experiment}
\label{5 Experiment}
In this section, we present experimental setup, comprehensive experimental results and detailed analysis, aiming at answering the following research questions (RQs):
\begin{itemize}

\item \textbf{RQ1:} Does  AOTree outperform the state-of-the-art methods in terms of recommendation accuracy?

\item \textbf{RQ2:} Does the considering aspect order obtained by AOTree help to improve  recommendation accuracy?

\item \textbf{RQ3:} Does the considering aspect order obtained from AOTree help to improve explainability?
\end{itemize}
\vspace{-0.8em}
\subsection{Experimental Setup}
\label{5_1 Experimental Setup}

\subsubsection{Dataset.}
\label{5_1_1 Dataset} The experiments are conducted on five publicly available user review datasets from Amazon 
and Yelp. 
For Amazon, we chose four common datasets: Cells Phones and Accessories, Office Product, Patio, Lawn and Garden and Digital Music. 

We first adopt the sentiment analysis toolkit~\cite{zhang2014users} and only keep users/items with more than $T_u/T_i$ review records due to the sparsity of the datasets~\cite{wang2018explainable, pan2022accurate}.
As we focus on the aspects extracted from reviews, we also set $T_a$ to the minimal number of appearing times for each aspect. 
The specific value of the threshold and the statistics of the datasets after preprocessing are shown in Table~\ref{tab:Summary of datasets}. The Density value denotes the average number of aspects mentioned by users. 
It is worth noting that the $T_u/T_i$ for Digital Music and Yelp are set differently in order to keep a relatively consistent density value. 
In the experiments, each dataset is divided into training set, validation set, and test set by 8:1:1. 

\begin{table*}[htb]\small
\vspace{-0.2cm}
\caption{Overall Performance on MSE on benchmark datasets. The best performance is in boldface. * and ** represent significant difference at 0.05 and 0.01 level, respectively, compared with the best baseline.}
\vspace{-0.3cm}
\label{tab:Overall Performance on MSE}
\centering
\begin{tabular}{c|c|c|c|c|c}
\hline
Model & Cells Phones and Accessories & Office Product & Patio, Lawn and Garden & Digital Music   & Yelp \\ \hline
MF      & 1.2184  & 0.8716  & 1.1089  & 1.1851  & 1.3994 \\
XGBoost & 1.0127  & 0.6980  & 1.0101  & 0.8589  & 1.2736 \\
NARRE   & 0.9967 & 0.6902  & 1.0292  & 0.8620  & 1.2535   \\
R3     & 0.9847  & 0.6882  & 0.9760  &  0.8278 & 1.2498 \\ 
ANR     & 0.9801  & 0.6810  & 1.0126  & 0.8189  & 1.2503 \\ 
TEM     & 0.9837  & 0.6771  & 0.9939  & 0.8247  & 1.2686 \\ 
ERRA     & 0.9811 & 0.7066  & 1.0270  & 0.8347  &  1.2364\\ \hline
AOTree  & \textbf{0.9719*}     &  \textbf{0.6729*}  & \textbf{0.9452**}  & \textbf{0.8050**} & \textbf{1.2140*}  \\ \hline
\end{tabular}%
\vspace{-2em}
\end{table*}
\begin{table*}[htb]\small
\vspace{-0.3cm}
\caption{Overall Performance on NDCG on benchmark datasets. The best performance is in boldface. * and ** represent significant difference at 0.05 and 0.01 level, respectively, compared with the best baseline.}
\vspace{-0.5em}
\label{tab:Overall Performance on NDCG}
\centering
\begin{tabular}{c|c|c|c|c|c}
\hline
Model & Cells Phones and Accessories & Office Product & Patio, Lawn and Garden & Digital Music   & Yelp \\ \hline
SULM      & 0.2576  & 0.3743  &  0.3456 & 0.3241  & 0.2363 \\
AOTree  & \textbf{0.2691*}     &  \textbf{0.3771*}  & \textbf{0.3467*}  & \textbf{0.3317*} & \textbf{0.2490*}  \\ \hline
\end{tabular}%
\end{table*}

\begin{table*}[htb]\small
\vspace{-0.3cm}
\caption{Effectiveness of aspect order on the AOTree model measured by MSE. $Basic$ means the original results by only using (Non-) Sensitive Users. $Shuffle$ and $Top@5$ denote the two disturbance operations applied to the aspect order.}
\vspace{-0.5em}
\label{tab:Explanation Effectiveness Performance}
\centering
\begin{tabular}{c|c|ccc|ccc}
\hline
\multirow{2}{*}{Dataset}     & \multirow{2}{*}{Overall} & \multicolumn{3}{c|}{Sensitive Users}    & \multicolumn{3}{c}{Non-Sensitive Users} \\ \cline{3-8} 
                             &                          & Basic       & Shuffle    & Top@5 & Basic       & Shuffle    & Top@5 \\ \hline
Cells Phones and Accessories &  0.9719 &  0.7355      &  0.7532   &    0.7628    &  1.2042    &  1.1708    &   1.1977    \\
Office Product             & 0.6729  & 0.5478 & 0.5763 & 0.5612  & 0.7924 & 0.7820  & 0.7901 \\
Patio, Lawn and Garden           &  0.9452 & 0.8973 &  0.9536  &  0.9446  &  1.04564  &  1.0156  &  1.0382  \\
Digital Music                & 0.8050 & 0.5888 & 0.6102 & 0.6379 & 0.8743 & 0.8691 & 0.8690 \\
Yelp                         &    1.2140 &  0.9774    &  0.9923     &  0.9879    &  1.4266    &    1.3234    &  1.3078    \\
\hline
\end{tabular}%

\end{table*}

\begin{table*}[htb]

\caption{Explainability of the Aspect Order on benchmark datasets. $Num\%$ denotes the coverage ratio of the reviews. $NDCG@5$ and $F1@5$ are the verification of the aspect order explanation only on the first 5 aspects. 
* and ** represent significant difference at 0.05 and 0.01 level, respectively, compared with the best baseline.}

\label{tab:Explainability of the Aspect Order}
\resizebox{\textwidth}{!}{
\begin{tabular}{c|ccc|ccc|ccc|ccc|ccc}
\hline
\multirow{3}{*}{Model} & \multicolumn{3}{c|}{Cells Phones and Accessories}                                                                & \multicolumn{3}{c|}{Office Product}                                        & \multicolumn{3}{c|}{Patio, Lawn and Garden}                                                    & \multicolumn{3}{c|}{Digital Music}                                                       & \multicolumn{3}{c}{Yelp}                                              \\ \cline{2-16} 
                         & \multicolumn{1}{c|}{Import} & \multicolumn{2}{c|}{Order}                             & \multicolumn{1}{c|}{Import} & \multicolumn{2}{c|}{Order}                             & \multicolumn{1}{c|}{Import} & \multicolumn{2}{c|}{Order}                             & \multicolumn{1}{c|}{Import} & \multicolumn{2}{c|}{Order}                             & \multicolumn{1}{c|}{Import} & \multicolumn{2}{c}{Order}                             \\ \cline{2-16} 
                         & \multicolumn{1}{c|}{Num\%}       & \multicolumn{1}{c}{NDCG@5} & \multicolumn{1}{c|}{F1@5} & \multicolumn{1}{c|}{Num\%}       & \multicolumn{1}{c}{NDCG@5} & \multicolumn{1}{c|}{F1@5} & \multicolumn{1}{c|}{Num\%}       & \multicolumn{1}{c}{NDCG@5} & \multicolumn{1}{c|}{F1@5} & \multicolumn{1}{c|}{Num\%}       & \multicolumn{1}{c}{NDCG@5} & \multicolumn{1}{c|}{F1@5} & \multicolumn{1}{c|}{Num\%}       & \multicolumn{1}{c}{NDCG@5} & \multicolumn{1}{c}{F1@5} \\ \hline
TEM                      & \multicolumn{1}{c|}{7.10}           &   \textbf{0.0158}     & 0.0101   & \multicolumn{1}{c|}{6.58}           &  0.0224      & 0.0522    & \multicolumn{1}{c|}{0.57}           &  0.0026      & 0.0016       & \multicolumn{1}{c|}{5.43}           &  \textbf{0.0560}     & 0.0329 & \multicolumn{1}{c|}{13.84}           & 0.0010   &  0.0580   \\
SULM                      & \multicolumn{1}{c|}{4.30}          &  0.0002      &  0.0031  & \multicolumn{1}{c|}{6.74}           &   0.0006     &  0.0076   & \multicolumn{1}{c|}{8.65}           &  0.0029      &  0.0097      & \multicolumn{1}{c|}{13.93}           &  0.0051     & 0.0210 & \multicolumn{1}{c|}{9.24}           &  0.0002  &  0.0020   \\ 
ERRA                      & \multicolumn{1}{c|}{10.61}          &  0.0099      &  \textbf{0.3257}  & \multicolumn{1}{c|}{13.87}           &  0.0167      & 0.2477    & \multicolumn{1}{c|}{8.48}           &     0.0018   &  0.2420     & \multicolumn{1}{c|}{11.30}           &   0.0013   & 0.2793 & \multicolumn{1}{c|}{19.06}           & 0.0010   & 0.2230    \\ \hline
AOTree                   & \multicolumn{1}{c|}{\textbf{10.87*}}           &      0.0013**          &     0.0587**           & \multicolumn{1}{c|}{ \textbf{24.04**}}           &   \textbf{0.0231*}            &   \textbf{0.2816**}         & \multicolumn{1}{c|}{\textbf{12.10**}}           &   \textbf{0.0037*}       &    \textbf{0.2578*}     & \multicolumn{1}{c|}{\textbf{33.62**}}           &   0.0025**              &    \textbf{0.3594**}      & \multicolumn{1}{c|}{\textbf{89.35**}}    & \textbf{0.0069**}  & \textbf{0.7170**}   \\ \hline
\end{tabular}
}

\end{table*}

\subsubsection{Compared Methods.}
\label{5_1_2 Baseline Methods}
In order to validate the performance of our proposed AOTree method, several baselines are selected for comparison. Specifically, we choose several commonly-used traditional and state-of-the-art methods in the field of recommender systems, such as the classic benchmark method (MF), tree-based methods (XGBoost and TEM), review-based method (NARRE, R3), and aspect-based prediction methods (SULM, ANR and ERRA). All the compared methods are evaluated based on the code published by the authors with careful hyper-parameter tuning. The compared methods are: 
\begin{itemize}

\item \textbf{MF}~\cite{koren2009matrix}: Matrix Factorization (MF) is a classic rating prediction method using bias terms and latent features for prediction. 

\item \textbf{XGBoost}~\cite{chen2016xgboost}: XGBoost is the state-of-the-art tree-based method. The final tree captures complex feature dependencies, that is the cross feature combination for different paths.

\item \textbf{TEM}~\cite{wang2018tem}: Tree-enhanced Embedding Method (TEM) is also a tree-based model. It is constructed into two cascade parts with GBDT and an easy-to-interpret attention network, making the recommendation process fully transparent and explainable.

\item \textbf{NARRE}~\cite{chen2018neural}: Neural Attentional Regression model with Review-level Explanations (NARRE) is a widely used state-of-the-art explainable recommendation method. This method focuses on the reviews usefulness and uses an attention mechanism to learn the importance weights over different reviews. 

\item \textbf{R3}~\cite{pan2022accurate}: Recommendation via Review Rationalization (R3) is a causal-aware explainable method, which extracts rationales from reviews via a rationale generator to alleviate the effects of spurious correlations in recommendation. Then, the final recommendation and causal-aware explanation can be generated according to the rationales.

\item \textbf{SULM}~\cite{bauman2017aspect}: Sentiment Utility Logistic Model (SULM) is a basic aspect-based model, which uses sentiment analysis of user reviews to identify the most valuable aspects for recommendation.

\item \textbf{ANR}~\cite{ANR}: Aspect-based Neural Recommender (ANR) is an end-to-end attention-based method, which performs aspect-based representation learning for both users and items via attention mechanism.




\item \textbf{ERRA}(~\citet{cheng2023explainable}): Explainable Recommendation by personalized Review retrieval and Aspect learning (ERRA) obtains recommendations and informative explanations by additional information obtained by retrieval enhancement.

\end{itemize}

\subsubsection{Evaluation Metrics.}
\label{5_1_3 Evaluation Metrics} 
The proposed AOTree method and all compared methods can be evaluated by the Mean Square Error (MSE), except SULM. 
For the MSE metric, described as Equation (\ref{mse}), a lower value means better performance. 
In order to make a comparison with the aspect-based algorithm, SULM, which converts the rating prediction problem into a preference classification problem for 0 and 1, we convert our scenario into a ranking task. 
To be more specific, we sort the recommended items into a list according to ratings for each user and evaluate the ranking performance by calculating the NDCG value. NDCG, short for Normalized Discounted Cumulative Gain, is a metric commonly used in recommendation systems to assess the quality of ranked recommendations~\cite{kanoulas2009empirical}. The formulas are as follows: 
\begin{small}
\begin{equation}
\label{ndcg}
\textstyle
    NDCG@K=\frac{DCG@K}{IDCK@K}
\end{equation}
\end{small}

\begin{small}
\begin{equation}
\label{dcg}
\textstyle
    DCG@K=\sum_{j=1}^K{\frac{2^{rel_j}-1}{log_2{(j+1)}}}
\end{equation}
\end{small}

\begin{small}
\begin{equation}
\label{idcg}
\textstyle
    IDCG@K=\sum_{j=1}^{K}{\frac{2^{rel_j}-1}{log_2{(j+1)}}}
\end{equation}
\end{small}

where $K$ denotes the position up to which the relevance is accumulated, and it is omitted and default to 5 in our method. $rel_j$ denotes the graded relevance at position $j$, and the set consisting of the top $K$ results is taken. Then, the final $NDCG$ is calculated by normalizing $DCG$ as Equation (\ref{ndcg}), and a larger value means better performance.

The goal of our method is to achieve human simulability, that is, to obtain considering aspect order as explanations. 
As the order of the aspects mentioned in reviews appears as a decision process, we could compare it with the predicted aspect order as the evaluation of the explanation. 
On the one hand, we verify the coverage of aspects, aiming to show the performance from the importance dimension ($Num\%$). 
On the other hand, inspired by the metrics in rank evaluation, we use $NDCG$ and $F1$ metrics to test the ordering effectiveness. 
Specifically, we assess the aspect composition order's quality at the aspect level using the $NDCG$ metric, and assess aspect coverage using the $F1$ metric.
According to the ``Order Effects Theory'' introduced above, the Primacy and Recency principles are crucial to the performance of orders. 
However, due to the fixed depth of the tree, we only test the front parts of the sequence to illustrate the Primacy principle, specifically considering the first five aspects of the sequence.

\subsubsection{Implementation Details.}
\label{5_1_4 Implementation Details}
For the compared algorithms, we follow the corresponding papers to initialize all the hyperparameters and carefully tune them to achieve optimal performance. 
We use Adaptive Moment Estimation (Adam)~\cite{kinga2015method} to optimize our method, with learning rate in \{0.00001, 0.0005, 0.001, 0005\}, L2 regularization coefficient in \{$10^{-3}$, $10^{-4}$, $10^{-5}$, $10^{-6}$, $10^{-7}$\} and dropout rate in \{0, 0.2, 0.4, 0.6, 0.8\}. 
We use early stopping with 10 epochs and report the final results from the best-performing model on validation sets. 
More specifically, other hyperparameters are searched with the batch size in \{16, 32, 64, 128\}, the latent factors number in the range [5, 20]. Further, in order to control the expense of constructing the AOTree, the max depth of the tree is limited to \{5, 10, 15,..., 100\} and the construction process does not require training.

\section{Results}

\subsection{Overview Performance (RQ1)}
\label{5_2 Overview Performance(RQ1)}

We first report the overall performance of AOTree in Table~\ref{tab:Overall Performance on MSE} and Table~\ref{tab:Overall Performance on NDCG}, which show the MSE and NDCG results according to the competitors. 
From the results, AOTree significantly outperforms all the other competitors on all datasets ($p<0.05$), which could answer RQ1. 
The performance of R3 method ranks second due to the dependence on causal relations rather than spurious correlations. And NARRE and ANR also show quite good performance, especially in Cell Phones and Accessories and Office Product datasets, which demonstrates that attention mechanism can capture implicit finer-grained properties from reviews. 
TEM captures the explicit cross features, which ignore the relationship from the ``ordering'' dimension, and it only shows a slight advantage in Office Product.
By exploring the ``ordering'' dimension among aspects, the AOTree is constructed by using sequence property and shows promising performance. 
From the ranking score results on NDCG compared with SULM, the AOTree method still significantly outperforms on all datasets ($p<0.05$), demonstrating its superiority in item ranking tasks.

Besides, we could also have several findings from the perspective of datasets. 
For the Patio, Lawn and Garden dataset, AOTree shows the best improvement over other methods, nearly $3.08\%$ ($p<0.01$) better than the second place method, TEM, which means most benefiting from the order effects phenomenon. 
In other words, the reviews written by users in this dataset well display such kinds of writing rules, and the reason may be that these users take the process of writing reviews more seriously, with a process of decision-making.

\subsection{Effectiveness of the Aspect Order (RQ2)}
\label{5_3 Effectiveness of the Aspect Order(RQ2)}

\begin{table*}[htb]\small

\caption{An example of aspect order generated by AOTree. The words in boldface in the original review are the extracted aspects by the sentiment analysis toolkit.}

\label{tab:case study}
{%
\centering
\begin{tabular}{l|p{0.85\linewidth}}
\hline
Explanation                           & \{\textbf{service}, lunch, location, tables, taste, \textbf{food}, staff, \textbf{prices}, flavor, \textbf{environment}, \textbf{service}, reservation, pizza\} \\ \hline
Original    & Great customer \textbf{service}, tasty \textbf{food} and drinks, and good \textbf{prices}.  What more do I need to say? In a city like Vancouver, where most dishes are overpriced, with entitled servers (expecting 20\% for barely passable \textbf{service}), and in a pretentious \textbf{environment}, you crave for something real: simply good \textbf{food}, \textbf{service}, and \textbf{prices}. That's it. And they hit all three notes, and that's how you create loyal customers who come back time and time again supporting your business. Thank you Suika, keep up the good work! \\
\hline
Groundtruth&\{\textbf{service}, \textbf{food}, \textbf{prices}, \textbf{service}, \textbf{environment}, \textbf{food}, \textbf{service}, \textbf{prices}\}\\ \hline
\end{tabular}%
}

\end{table*}

\begin{table}[t]

\caption{Ablation study on MSE for AOTree.}

\label{tab:Ablation Study}
\centering
\resizebox{\columnwidth}{!}
{
\begin{tabular}{c|c|c|c}
\hline
   & Cells Phones \& Accessories & Patio, Lawn \& Garden & Digital Music  \\ \hline
w/o Tree  & 1.0130  &  0.9869  & 0.8135   \\
w/o PEL    &  1.0175  &  0.9814  & 0.8130  \\
w/o SAL    & 1.0116   &  1.0642  & 0.8220   \\
w/o LN   & 1.0179   &  0.9946  & 0.8191   \\ \hline
AOTree   & \textbf{0.9719} &\textbf{0.9452}  & \textbf{0.8050}       \\ \hline
\end{tabular}%
}

\end{table}

For RQ2, we perturb the obtained aspect order to demonstrate the superiority of the current order.
However, the results are not stable for the whole dataset. So we split users according to the MSE results shown in Table~\ref{tab:Overall Performance on MSE}. To be specific, we focus on the samples with MSE less than the average value in the training set. If all interactions for one user meet the above criterion, the user will be selected. In fact, this is a very strict partition criterion since it requires each of a user's interactions meets such a criterion, and the selected users should have strong order effect. So we mark the selected users as \textit{Identified Strong Sensitive Users} and the others as \textit{Identified non-Strong Sensitive Users}. Our analysis suggests that the fraction of \textit{Identified Strong Sensitive Users} reaches $20\%$ to $30\%$ in different datasets ($20.37\%$ in Cells Phones and Accessories, $26.61\%$ in Office Product, $21.60\%$ in Patio, Lawn and Garden, $21.05\%$ in Digital Music, and $31.57\%$ in Yelp). Such a proportion is close to that of the \textit{Strong Sensitive Users} with $consistency\_dis_{user}$ higher than 0.5 in our Preliminary Analysis, indicating our method's effectiveness in capturing order effect.


Then, for each user, we verify the effectiveness of the captured aspect order for each interaction by adopting two perturbation operations related to the ``ordering'': 1) shuffling the whole order to verify whether the sequence generated by the model is effective, and 2) replacing the first five aspects with random ones to see whether the Primacy principle contributes. 
The final results are shown in Table~\ref{tab:Explanation Effectiveness Performance}. 
From the results, we can obtain the following findings. For the \textit{Identified Strong Sensitive Users}, the results after perturbation show higher MSE values, which suggests the superiority of the aspect order obtained by our method. While for the \textit{Identified non-Strong Sensitive Users}, the above conclusion seems to be dismissed since the results after perturbation do not make a significant difference. 



\subsection{Explainability of Aspect Order (RQ3)}
\label{5_4 Explainability of the Aspect Order(RQ3)}

As we seek for a specific form of interpretability known as human simulability, we finally verify the explainability of the AOTree model according to the simulability, that is, to attain the matching degree between the obtained aspect order and the original aspect sequence that appeared in the reviews. 
We compare the importance and order of the first five aspects to prove that our method could predict users' decision process displayed as reviews. 
The importance is represented by calculating the coverage of the aspects, and the order is verified by numerical metrics of NDCG and F1. 

We choose TEM, SULM and ERRA as the compared methods in the explanability evaluation because they also focus on using the extracted aspects as explanations. 
The aspect order of TEM and SULM is constructed according to the rank of the attention weights for each aspect and the aspect order of ERRA is extracted by the aspects appeared in the generated explanations.
The results are shown in Table~\ref{tab:Explainability of the Aspect Order}. For the $Num\%$ value, we can see that AOTree significantly outperforms the other three methods on all the datasets, which means that our method has the advantage of more accurately discovering the considering factors in the reviews. 
The advantage is quite significant, especially for the Yelp dataset ($p<0.01$). 
And for the $NDCG@5$ and $F1@5$ metrics, although TEM performs better on Cells Phones and Accessories and Digital Music dataset, the coverage value is quite low (only 7.10\% and 5.43\%). The reason for this abnormal result is that in these two datasets, the proportion of sensitive users is the lowest (20.37\% in Cells Phones and Accessories and 21.05\% in Digital Music), resulting in the order effects generated by the model not fully matching the results in actual user reviews. Also, ERRA performs relatively good results on Cells Phones and Accessories, which show its advatange on aspect prediction due to the aspect enhancement component.
In summary, we can conclude that AOTree outperforms TEM and SULM in terms of explainability.

\subsection{Case Study}

\label{5_5 Case Study}

We use a case study (from a randomly picked review) to illustrate the explanations represented by our method. 
The results are shown in Table~\ref{tab:case study}. 
Considering the context of recommending a restaurant to a user, our model presents the explanations of why the restaurant is recommended as an aspect sequence shown in the first row of the table.
The original review is also shown with extracted aspects in boldface, and the ground-truth aspects are shown in the last row.
From this case study, we can see that AOTree can find that the user may be concerned with the aspect of \textbf{service}, \textbf{lunch}, \textbf{location} and so on for this item, which follows the specific decisive order, leading to the final decision-making behavior.
The order of the aspects can mimic the decision process as reviews, which is mostly consistent with the ground-truth order.
This qualitative study further confirms the effectiveness of the proposed AOTree method.

\subsection{Ablation and Hyper-parameter Study}
\label{5_6 Ablation and Hyper-parameter Study}

\subsubsection{Ablation Study}
\label{5_6_1 Ablation Study}

The ablation study for each designed component in AOTree is considered to justify its validity. 
As our main goal is for accuracy, we conduct ablation experiments on the MSE metrics.
The two main components are specifically the AOTree Generator (Tree) and the Prediction Generator. 
Furthermore, we split the Prediction Generator into more detailed parts, which are the Position Embedding Layer (PEL), the Self-Attention Layer (SAL) and the Layer Normalization (LN). 
We test it on three datasets, and the MSE results are shown in Table~\ref{tab:Ablation Study}, where $w/o$ means the MSE results without the corresponding component. 
When replacing each of the components, the corresponding MSE value increases, especially for SAL part and LN part.
Although the design of AOTree is not complicated and is constructed by several independent components, the ablation results could confirm the effectiveness of each design.

\subsubsection{Hyper-parameter Study}
\label{5_6_2 Hyper-parameter Study}

\begin{figure}[t]
    \centering

\includegraphics[width=0.999\linewidth]{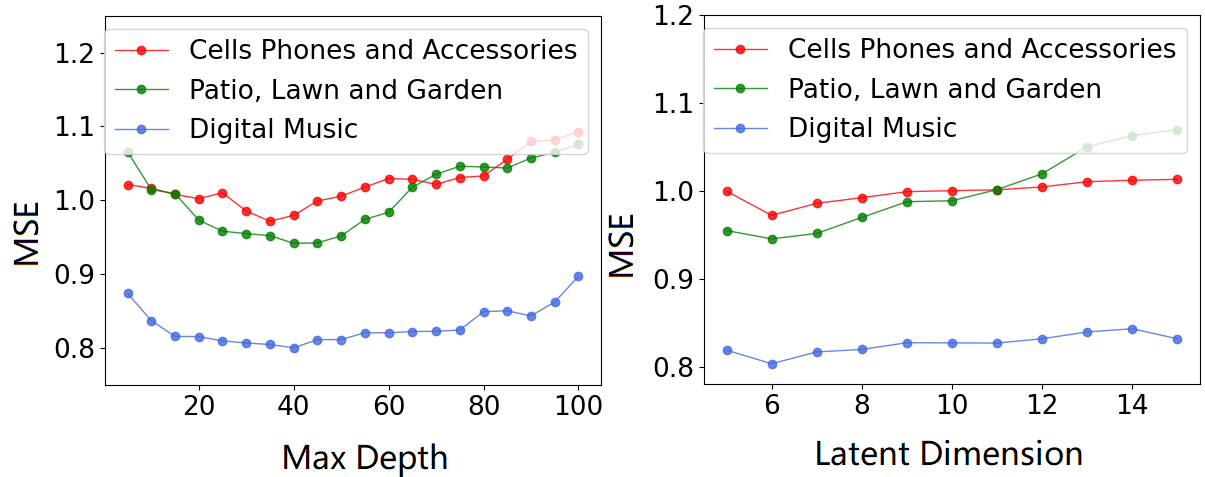}

    \caption{Hyperparameter analysis w.r.t. the max depth and latent dimension size on three datasets.}

\label{fig:para}
\end{figure}

We empirically analyze the impacts of two key hyper-parameters: the max depth of the tree and the size of the latent dimensions. The results are shown in Figure~\ref{fig:para}. 
For the max depth of the tree, it limits the length of the aspect order, which is related to the users' considering process. 
The figure presents the performance w.r.t. MSE loss over the varying number of the max depth. 
Although the optimal max depth values are not the same for different datasets, they show a similar trend. 
With increasing max depth, the MSE loss of AOTree gradually decreases at first and then increases afterward. 
Generally, when the value of the max depth is set at the range of $[35, 45]$, we can observe the best performance empirically. 
Then, we consider the empirical value of the latent dimension.
The embedding component is related to the user/item/aspect id, which shows the ability to represent each variable. 
As shown in the figure, the best performance is achieved when the latent dimension is set at the range between $[5, 10]$. 

\subsection{Time Complexity Analysis}
\label{4_6 Time Complexity Analysis}

We separate complexity analysis into two parts since there are two phases in the learning process of our method. For the AOTree building phase, the time complexity is $O(m \cdot e \cdot l \cdot \log l)$ or $O(n \cdot e \cdot l \cdot \log l)$, where $m$/$n$ represents the number of users/items for User-AOTree/Item-AOTree, $e$ is the maximum depth of trees, and $l$ is the number of extracted aspects. For the embedding phase, the time complexity is $O(2 \cdot a \cdot d \cdot N)$, where $a$ is the attention size, $d$ is the embedding size and $N$ is the number of training instances. Above all, the overall time complexity of our model is $O(l \cdot e \cdot m \cdot \log l + l \cdot e \cdot n \cdot \log l + 2 \cdot a \cdot d \cdot N)$.
The time complexity of the comparison algorithm TEM is $O(S\cdot e \cdot \Vert x \Vert _0 \cdot logN + 2\cdot a\cdot d\cdot S\cdot N)$, where $S$ and $\Vert x \Vert _0$ denote the number of trees and the average number of non-zero entries in the training instances, respectively. Although our method introduces the aspect-level complexity ($l$), we simplify the procedure of feature fusion from multiple trees, thereby reducing the overall complexity and making it acceptable.

\section{Discussion}

In this work, we confirm both the existence and variance of order effects within user reviews and design a new model named AOTree to capture the order effects. Extensive experiments exhibit AOTree's multiple strengths like more accurate rating predictions and better aspect order explanations compared with the related commonly-used traditional methods and state-of-the-art methods in the field of recommender systems. However, we identify that aspect order mining and modeling is essentially a challenging task and our model still has several limitations. The following discusses the limitations in our work and the potential problems of AOTree, shedding light on the future investigation and improvements.

Our preliminary analysis suggests that although most users (nearly 70\%) have specific aspect order considerations and most items (over 60\%) tend to have a specific aspect order considered by users, the order effect seems strong for just a small proportion of users and items. For example, only about 20\% users’ $consistency\_dis_{user}$ values are higher than 0.5, and approximately 20\% items' $consistency\_dis_{item}$ values are higher than 0.5, making mining aspect order for users and items complicated and challenging. As a result, the proportion of \textit{Identified Strong Sensitive Users} uncovered in our experiments is about 20\% to 30\% in different datasets. We thought data sparsity is the primary reason for the above problem, i.e., $73\%$ users only contribute less than 5 reviews, providing limited corpus for aspect mining. Data sparsity is an essential problem in recommender systems, and there have been abundant advanced techniques like auxiliary information integration~\cite{cheng2023explainable, zheng2019deep}. 
The future research can investigate how to combine these techniques with our model to strengthen the effectiveness of aspect order mining, especially for users and items with limited reviews.

While our method provides promising solutions, it might aggravate the problems embedded in recommender systems. Our work focuses on utilizing users' decision aspect order preference for recommendations, which may inadvertently exacerbate filter bubbles by reinforcing users' existing preferences on aspect order, potentially leading to less exposure to diverse perspectives and aggravating polarization. To mitigate such risks, further research can explore to integrate diversity-enhancing mechanisms (e.g., incorporating some new items) when leveraging a user's aspect order preference, promoting exposure to different perspectives.


\section{Conclusion}
\label{7 Conclusion}

This work focuses on the Order Effects Theory of aspects in explainable recommendation and seeks a specific form of interpretability, known as human simulability.
We first verify the applicability of Order Effects Theory by analyzing the Yelp dataset.
Then, by applying the theory, we proposed Aspect Order Tree-based (AOTree) explainable recommendation, which is achieved to capture the dependency relationships among decisive factors. Experiment results show that the proposed method can perform well on public datasets, achieving higher accuracy as well as better interpretability according to particular forms.


\section{Ethical Statement}

The study was conducted on commonly used public datasets. All the data is anonymous and allowed to use. The result of our research conducted positive outcome for effective recommendation. In our perspective, our work does not exists negative society impacts or potential misuse, also, the data collection process and modeling process comply with the procedure without ethical implications.

\small
\bibliography{arxiv}




\end{document}